\documentclass[12pt]{article}

\usepackage{amsfonts}
\usepackage[margin = 2.7 cm]{geometry}
\usepackage{amsmath,amssymb}
\usepackage{amsthm}
\usepackage{colortbl}
\usepackage{color}
\usepackage{xcolor}
\usepackage{listings}
\usepackage{url}
\usepackage{setspace}
\usepackage{csquotes}
\usepackage{tgpagella}
\usepackage{lscape}
\usepackage[italic]{mathastext}
\usepackage{enumitem}
\usepackage{isomath}
\usepackage{mathtools}

\usepackage{booktabs}
\usepackage{tabularx}
\usepackage{multirow}
\usepackage{adjustbox}

\usepackage[mode=buildnew]{standalone}

\usepackage{hyperref}
\usepackage[backend=biber, style=authoryear, maxbibnames=10, maxcitenames=2, minbibnames=10 ,uniquelist=minyear, natbib]{biblatex} 
\makeatletter
\DeclareCiteCommand{\citeyear}
  {}
  {\printtext[bibhyperref]{\printfield{year}\printfield{extrayear}}}
  {\multicitedelim}
  {}
\makeatother

\usepackage{graphicx}
\usepackage{subcaption}
\usepackage[font=small,labelfont=bf, width=\textwidth]{caption}

\renewbibmacro*{doi+eprint+url}{%
\iftoggle{bbx:url}
  {\iffieldundef{doi}{\usebibmacro{url+urldate}}{}}     
  {}%
\newunit\newblock
\iftoggle{bbx:eprint}
  {\usebibmacro{eprint}}     
  {}%
\newunit\newblock
\iftoggle{bbx:doi}     
  {\printfield{doi}}     
  {}}

\AtEveryBibitem{\clearfield{issn}}
\AtEveryCitekey{\clearfield{issn}}

\AtEveryBibitem{\clearfield{month}}
\AtEveryCitekey{\clearfield{month}}

\AtEveryBibitem{\clearlist{language}}
\AtEveryCitekey{\clearlist{language}}

\usepackage{listings}
\usepackage{comment}
\hypersetup{
	colorlinks   = true, 
	urlcolor     = blue, 
	linkcolor    = black, 
	citecolor   = red 
}

\usepackage{tocloft}
\cftsetindents{section}{1em}{1.5em}

\usepackage{algorithm} 
\usepackage{algpseudocode}
\algrenewcommand\algorithmicrequire{\textbf{Input:}}
\algnewcommand\algorithmicforeach{\textbf{for each}}
\algdef{S}[FOR]{ForEach}[1]{\algorithmicforeach\ #1\ \algorithmicdo}

\newtheoremstyle{thmstyle}
  {\topsep} 
  {\topsep} 
  {} 
  {} 
  {\bfseries} 
  {.} 
  {.5em} 
  {} 
\theoremstyle{thmstyle}

\usepackage{parskip}

\usepackage{bbm}  
\newcommand{\1}{\ensuremath \mathbbm{1}}

\author{Yves Achdou\footnote{Laboratoire Jacques-Louis Lions (LJLL), University Paris Cité, achdou@ljll.univ-paris-diderot.fr} \and Johannes Brumm\footnote{Karlsruhe Institute of Technology, johannes.brumm@kit.edu} \and Lukas Frank\footnote{Karlsruhe Institute of Technology and LJLL, Université Paris Cité, lukas.frank@kit.edu}  }

\title{Mastering Stochastic OLG Models \\ in Continuous Time\thanks{We thank
participants of the 32nd International Conference on Computational Economics and Finance (Venice, 2026), the workshop on mean field games and mean field control in economics (Paris, 2025), the conference on machine learning for economics and finance (Torino, 2025), the doctoral seminar UMA at ENSTA (Paris, 2026), and of the seminar at KIT (Karlsruhe, 2025) for helpful comments. We acknowledge financial support from the ERC project SOLG for Policy (101042908). Moreover, we acknowledge support by the state of Baden-Württemberg through bwHPC
and the German Research Foundation (DFG) through grant INST 35/1597-1 FUGG.}}

\date{ 
\today \vspace{0.5cm} \\ --- \textit{Preliminary and Incomplete} ---}

\begin{document}

\newcommand{\E}{\ensuremath \mathbb{E}}
\newcommand{\R}{\ensuremath \mathbb{R}}
\newcommand{\V}{\ensuremath \mathcal{V}}
\newcommand{\G}{\ensuremath \mathcal{G}}

\maketitle

\begin{abstract}
\noindent We propose a comprehensive framework for solving overlapping-generations (OLG) models in continuous time with both idiosyncratic and aggregate risk. Our general characterization of equilibrium through the master equation operates on the joint distribution over the continuous idiosyncratic states, age and wealth. Our computational strategy is to take a finite-dimensional representation of this distribution as an input of a neural net which in turn outputs a finite-difference representation of the (conditional) value function. This idea can be applied generally to heterogeneous agent models with aggregate risk, and we call it \textit{finite-difference neural operator}. 
Our method combines advantages from modern neural nets and traditional finite-difference methods: It is grid-free in the high-dimensional distribution, and retains control on boundary conditions in low-dimensional state variables. 
Moreover, our method is able to enforce shape constraints. We showcase its flexibility by solving a continuous-time OLG model with aggregate risk alone where we characterize the distribution by its supporting function; and to an OLG model with both types of risk.
\end{abstract}

\newpage

\section{Introduction} \label{sec:Intro}

Overlapping generations (OLG) models have been used to study intergenerational welfare effects of public policy since the seminal work of  \citet{samuelson1958} and \citet{diamond1965}. In this model class, agents are necessarily heterogeneous, at least in their age. Further drivers of heterogeneity, such as idiosyncratic risk, might enter the model on top of that. At the same time, one often wishes to include sources of aggregate risk, for instance to properly account for risk-sharing effects of public policy. However, solving rational expectations models with heterogeneous agents and aggregate risk is notoriously difficult as it requires agents to keep track of the entire distribution to forecast prices.\footnote{There have been several attempts of avoiding this challenge by relaxing the rationality assumption, see, e.g., \citet{sargent1993}, \citet{kubler2025b}, and \citet{moll2025}.} This poses a formidable numerical challenge.


To address the computational challenge in OLG models with aggregate risk, (1) we cast these models in continuous time and characterize equilibrium as the solution to a master equation that operates on the joint distribution over idiosyncratic states, (2) we project the infinite-dimensional distribution on a finite-dimensional representation, and (3) we propose a novel approach, the \textit{finite-difference neural operator}, to efficiently solve the master equation using that representation.

Recasting the OLG model in continuous-time allows to leverage recent advances from the literature on mean-field games and to characterize equilibrium by means of a so-called master equation (\cite{cardaliaguet2019}). This master equation features a derivative term with respect to the infinite-dimensional distribution of agents over their idiosyncratic states. Said derivative term is the main difficulty in solving the model. In a model without idiosyncratic risk, we can exploit the fact that the distribution of agents of a given age becomes singular: all agents of one generation are equal. This enables us to represent the distribution of agents as a one-dimensional function, the \textit{generational wealth function}. In a model with idiosyncratic risk, the distribution is truly non-singular; an additional challenge that we embrace to showcase the flexibility of our method.

In the next step, we project the distribution -- or the generational wealth function in the case with aggregate risk alone -- on a finite but relatively high-dimensional space. Therefore the derivative of the value function with respect to the distribution reduces to computing a gradient on a high-dimensional space. Our approach is, in principle, agnostic to the projection method. In practice, the chosen projection must at the same time be fast and still reflect the features of the distribution that matter for the application at hand.

To solve the model with a finite-dimensional representation of the distribution, crucially, we construct a neural net that takes this distribution as an input and outputs a finite-difference discretization of the value function conditional on the given distribution. This approach is a natural extension of the finite-difference solution to a model without aggregate risk as presented, e.g., in \citet{achdou2022}, and we refer to it as \textit{finite-difference neural operator}. Importantly, it allows to control boundary conditions and to enforce shape-constraints. For instance, we can treat Dirichlet conditions on terminal ages and a state-constraint at zero wealth (the borrowing constraint) in exactly the same way as in the model without aggregate risk.
Additionally, we enforce concavity in the wealth dimension by learning second partial derivatives of the value function instead of plain values. By constraining these second derivatives to negative values, we obtain, after integration, a concave value function.
Neural nets are an appropriate choice as they excel in high dimensions and allow for fast computations of the high-dimensional gradient with respect to the discretized distribution by means of automatic differentiation. Hence, we can efficiently evaluate the master equation, i.e., the equilibrium conditions. We then use a gradient-descent method to minimize violations of these equilibrium conditions in the spirit of \citet{maliar2021} and \citet{azinovic2022}.

We apply our finite-difference neural operator to a challenging overlapping-generations model with substantial aggregate (and idiosyncratic) risk. Aggregate risk takes the form of shocks to total factor productivity and depreciation. Idiosyncratic risk, on the other hand, manifests as shocks to labor income.
We equip the continuous-time OLG model with a sufficiently rich structure to showcase the capabilities of our method: a stochastic death process, a pay-as-you-go social security system, and a bequest motive à la \citet{denardi2004}. We consider two applications: A model with aggregate risk alone -- where the sole source of heterogeneity is ageing -- as well as a model with aggregate and idiosyncratic risk, with heterogeneity along age and labor income.

Our finite-difference neural operator solves the OLG model with substantial aggregate (and idiosyncratic) risk to satisfactory levels of accuracy. In particular, thanks to the finite-difference representation, we are able to capture individual dynamics around the borrowing constraint well.\footnote{Accounting for the borrowing constraint is a considerable challenge, as highlighted by \citet{gu2024}.} All experiments converge in less than 24 hours on a state-of-the-art GPU (Nvidia H200).

\paragraph{Related Literature}

This paper is part of the growing literature on global solution methods for high-dimensional economic models.
%
%
Early contributions to the literature are grid-based and mitigate the curse of dimensionality through (adaptive) sparse grids \citep{krueger2004, judd2014, brumm2017, schaab2022}.
%
%
More recent grid-free methods include Gaussian processes \citep{scheidegger2019, eftekhari2022} and tensor-train approximations \citep{brumm2026}. However, the majority of the literature focuses on techniques from deep learning. After a first wave of pioneering work has provided solid proof-of-concepts \citep{maliar2021, azinovic2022, ebrahimikahou2021, han2022}, the ongoing second wave of research is demonstrating broad applicability and explores refinements in the methodology \citep{kase2024, pascal2024, kahou2024,  gu2024, azinovic2024, folini2025, fernandez-villaverde2025, gopalakrishna2026, gopalakrishna2026a, yang2026, kase2026, payne2026, zhong2026, azinovic-yang2026, druedahl2026}.

Among this second wave, \citet{gu2024} is particularly important for our research as it is, to the best of our knowledge, the first attempt to approximate solution to a master equation in a continuous-time Krusell-Smith model. We bring this continuous-time master equation approach to the class of OLG models and advance on several challenges emphasized by \citet{gu2024}, i.e., ensuring crucial shape properties of the value function and dealing with a hard borrowing-constraint.

Regarding our computational methodology, the  closest papers are \citet{zhong2026, azinovic-yang2026} that both apply neural operators, although in quite different versions. Like \citet{zhong2026}, we take discretized distributions as an input, while \citet{azinovic-yang2026} use a truncated sequence of shock histories. Like \citet{azinovic-yang2026}, our neural operator yields coefficients of a simple grid function, whereas \citet{zhong2026} remains more faithful to the original work of \citet{li2021} and learns mappings in Fourier space. Similar to \citet{azinovic-yang2026}, we leverage the grid-representation for shape-preservation of the approximated value function. Beyond this, we also exploit it to employ an upwind scheme and to enforce boundary conditions.

From a model perspective, to the best of our knowledge, we are the first to solve a continuous-time stochastic OLG model with aggregate risk. In discrete time, the inclusion of aggregate risk has been studied by, among others, \citet{krueger2006, brumm2017b, harenberg2019, azinovic2022, , brumm2025, azinovic-yang2026}. The few existing continuous-time OLG studies feature idiosyncratic risk only \citep{barczyk2018, schaab2022, chen2026}.


\paragraph{Contents.}

The remainder of this paper is organized as follows. 
Section~\ref{sec:GeneralSolutionMethod} describes our general solution method for heterogeneous-agent models with aggregate risk. Section~\ref{sec:Model} presents an OLG model with aggregate and idiosyncratic risk in continuous time. In section~\ref{sec:SolvingAggRisk}, we show how to apply our general method to solve a version of said OLG model where idiosyncratic risk is turned off. Section~\ref{sec:SolvingAggIdioRisk} treats the case with both risks, i.e. aggregate and idiosyncratic risk. 

\clearpage

\section{A general method for HA models with aggregate risk} \label{sec:GeneralSolutionMethod}

In this section, we describe in an abstract way the general idea of our solution method, the finite-difference neural operator. Notation in this section is independent from the other sections.

Let $X = X_1 \times X_2$ be a vector space. In the applications below, $X$ will be the state space, with $X_1$ being a low-dimensional component (e.g., the idiosyncratic state variables), and $X_2$ a relatively high-dimensional approximation to an infinite-dimensional distribution.
Hence,  $X_1 \cong \R^{q_1}$ and $X_2 \cong \R^{q_2}$, with $q_2 \gg q_1$.  Our goal is to approximate a value function $V: X_1 \times X_2 \rightarrow \R$ that solves a nonlinear, first-order PDE
$$F(x_1, x_2, V, \partial_{x_1} V, \partial_{x_2} V) = 0, \qquad (x_1, x_2) \in E,$$
on a subset $E\subset X$. In practice, this subset may be, for instance, the ergodic set.

Before defining our solution method, let us reinterpret the task. We denote by $\mathbb{R}^{Y}$ the space of functions from $Y$ to $\mathbb{R}$. Instead of approximating a \emph{function} $V\in \mathbb{R}^{X_1 \times X_2}$, we search for an \emph{operator} $\V: X_2 \rightarrow \mathbb{R}^{X_1}$ and define 
$$V(x_1, x_2) \coloneqq \V(x_2)(x_1).$$ 

We thus approximate the value function in two steps: 
First, approximate the elements of $\mathbb{R}^{X_1}$ by a family of functions~$G(\cdot | \theta_1): X_1 \rightarrow \mathbb{R}$, with parameters $\theta_1 \in \mathbb{R}^M$. We denote the set of these functions by $\mathcal{G}$. 
Second, approximate the discretized value operator $\widetilde{\V}: X_2 \rightarrow \mathcal{G}$ with the help of a function $f(\cdot|{\theta_2}): X_2 \rightarrow \mathbb{R}^M$. We also write, in short, $f_{\theta_2}$. This function is parameterized by $\theta_2\in\mathbb{R}^N$, takes high-dimensional inputs $x_2\in X_2$, and outputs parameter values $\theta_1$ for $G$ conditional on $x_2$.
The overall approximation of the value function is given by $$\widetilde{V}_{\theta_2}(x_1, x_2) \coloneqq G\big(x_1 \big| {} f_{\theta_2}(x_2)\big) \approx V(x_1, x_2).$$

More specifically, for the approximation $G(\cdot | \theta_1)$, we use a $q_1$-dimensional grid with $M$ grid points.\footnote{In general, $M$ scales badly with dimension $q_1$. For instance, with a uniform grid and $m$ points per dimension, one has $M=m^{q_1}$. Sparse grids \citep{krueger2004} mitigate this curse of dimensionality, and adaptive sparse grids \citep{brumm2017} even more so. In principle, our approach can easily be combined with a sparse grid on $X_1$, not so trivially with an adaptive one, though.} This offers two advantages: We keep tight control on the behavior at boundaries of $X_1$ and we can additionally enforce shape-constraints or other qualitative properties of the solution, as described in the subsequent sections. For the latter approximation, $f_{\theta_2}$, we use a neural net with $N$ parameters because they are known to cope well with high dimensions \citep{berner2022}. In particular, we can compute partial derivatives $\partial_{x_2}f_{\theta_2}$ by means of automatic differentiation.\footnote{\label{fn:GeneralMethod_inputs_not_too_large}Our neural operator has, typically, a much larger output size than input size, $M>q_1$. We therefore compute partial derivatives $\partial_{x_2}f_{\theta_2}$ in forward mode with complexity $\mathcal{O}(q_1 \cdot \text{cost}(f_{\theta_2}))$ as opposed to reverse-mode automatic differentiation with complexity $\mathcal{O}(M \cdot \text{cost}(f_{\theta_2}))$, where $\text{cost}(f_{\theta_2})$ designates the number of operations required to evaluate $f_{\theta_2}$ \citep{baydin2015}. This means, however, that we need to keep $q_1$ of moderate size. In our applications, $q_1$ is not larger than a few hundreds -- a number we find sufficient to approximate the distributions at hand.}

In total, our method is specified by choosing a grid $G$ of size $M$ and defining a neural net architecture with $N$ parameters, that maps a $q_2$-dimensional input vector of high-dimensional state variables on an $M$-dimensional output vector of grid values in the low-dimensional state space components. 

To learn the free parameters $\theta_2\in\mathbb{R}^{N}$, we use a stochastic gradient-descent method minimizing the squared PDE residuals $|F(x_1, x_2, \widetilde{V}_{\theta_2}, \partial_{x_1} \widetilde{V}_{\theta_2}, \partial_{x_2} \widetilde{V}_{\theta_2})-0|^2$ on a finite approximation $\widetilde{E}$ of $E$. We refer to this method as \textit{finite-difference neural operator}, since $f_{\theta_2}$ constitutes a neural operator\footnote{Neural operators have been pioneered in the deep learning community by \citet{lu2021} and \citet{li2021}.}, and, in practice, we use upwind finite differences on grid $G$ to compute $\partial_{x_1} \widetilde{V}_{\theta_2}$.

\begin{figure}
    \centering
    \includegraphics[width=0.8\textwidth]{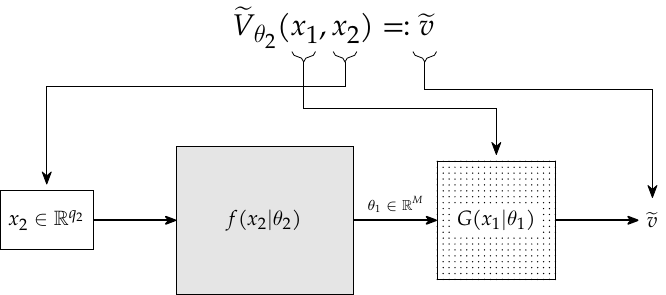}
    \caption{General finite-difference neural operator. It takes the high-dimensional variable $x_2$ as an input and outputs the $M$ values on a low-dimensional grid in $X_1$. We obtain the evaluation of $x_1 \mapsto \widetilde{V}_{\theta_2}(\cdot, x_2)$ by interpolation on said grid.}
    \label{fig:SolutionMethodGeneral}
\end{figure}

\clearpage

\section{Model} \label{sec:Model}
\subsection{General model description}

\paragraph{Demographics.}

Time is continuous, $t\geq 0$. At each point in time $t$, a new generation of mass $G$ is \enquote{born to the economy} at age $a=\underline{a}$, consisting of 
infinitely many heterogeneous agents.
Time to death $\tau$ follows a Poisson process with inhomogeneous death rate $\pi(a)$. All surviving households die at maximum age $\overline{a}$.

We denote the exogenous demographic measure by $\lambda_t\in \mathcal{M}(A)$.\footnote{For every set $Y$, let $\mathcal{M}(Y)$ be the set of measures on the Borel $\sigma$-algebra of $Y$.} This measure depends only on generation sizes at birth and on death probabilities,
\begin{equation}\label{eq:distribution_lambda}
    \lambda_t(\cdot) = \int_\cdot \underbrace{G e^{-\int_{\underline{a}}^a\pi(b)db}}_{\eqqcolon f_t^{\lambda}(a)}\, da.
\end{equation}
We denote its Lebesgue density by $f_t^{\lambda}$ and call it  demographic density.

\paragraph{Risk.} The economy is subject to aggregate risk via a stationary continuous-time Markov chain $(Z_t)$ with finite states $\mathcal{Z} = \{z_1, \dots, z_p\}$ and $Q${}-matrix $Q^{Z}\in\mathbb{R}^{p\times p}$.\footnote{We adopt the definition of continuous-time Markov chains as given in, e.g., \citet{chung1960}.} Let the Markov chain have a stationary distribution which we denote by $\eta_Z$. The stochastic variable $Z_t$ affects the economy through the production process, see below. 

Additionally, households face idiosyncratic risk. Consider a household $i$ from generation born at time $s$. Their labor productivity follows a stationary continuous-time Markov chain $(\varepsilon^{si}_t)$ with finite states $\mathcal{E} = \{e_1, \dots, e_q\}$ and $Q${}-matrix $Q^\varepsilon\in\mathbb{R}^{q\times q}$. Let the Markov chain have a stationary distribution which we denote by $\eta_{\varepsilon}$. Processes $(\varepsilon^{si}_t)$ are iid within generation $s$ for all $i$, independent from the individual time-to-death probabilities $\tau$, and independent from $(Z_t)$. We drop superscripts $si$ when there is no risk of ambiguity.

\paragraph{Households.}

Households supply labor according to an exogenous labor profile $\ell$, earn wage $w$, and either spend their income on consumption, $c$, or accumulate savings to a stock of assets, $x$. Assets yield interest at rate $r$ and are subject to a borrowing constraint, $x\geq 0$. Households stop working when they reach retirement age $a_{\text{ret}}$. We equip households with CRRA preferences on consumption and non-homothetic preferences on bequests à la \citet{denardi2004}. That is, consumption preferences are given by
\begin{equation} \label{eq:CRRA_utility}
    u(c) = \frac{c^{1-\sigma}-1}{1-\sigma}, \qquad \sigma > 0, \; \sigma \neq 1,
\end{equation}
with risk aversion parameter $\sigma$. The bequest motive is
\begin{equation} \label{eq:bequest_utility}
    U(B) = \phi_1 \big(1+\frac{B}{\phi_2}\big)^{1-\sigma},
\end{equation}
with bequest desirability parameter $\phi_1$ and luxury good parameter $\phi_2$.

The economy features a standard pay-as-you-go (PAYG) social security system with defined contributions: Young households, $a<a_{\text{ret}}$, pay a fixed share $\zeta$ on their gross labor income, $w\varepsilon\ell(a)$, to finance the retirement benefits $\varsigma$ of the old, $a \geq a_{\text{ret}}$. Retirement benefits are thus simply given as contributions divided by the mass of  beneficiaries

\begin{equation}
    \varsigma = \frac{\int_{\underline{a}}^{a_{\text{ret}}} \zeta w \varepsilon \ell(a) \,\eta_{\varepsilon}(d\varepsilon) \, \lambda(da)}{\lambda([a_{\text{ret}}, \overline{a}])}.
\end{equation}

When dying, households leave bequests. We model parental links between generations with an exogenous \enquote{parent-age} probability distribution $p(a)$ over age when giving birth. We collect the bequests left per generation, $b^{\text{out}}(a)$, and distribute them equally among their children, $b^{\text{in}}(a)$.\footnote{When households die before their children are economically active, they bequeath to their \enquote{siblings} from the same generations. This assumption is quantitatively not important since few households die young.} A precise formula for this procedure will be given in equations~\eqref{eq:model_b_out}-\eqref{eq:model_b_in}.

We denote by $\iota$ the total net income of households, i.e. capital income, net labor income, retirement benefits, and received bequests. It is given by
\begin{align}
    \iota(a, x, \varepsilon, r, w, b) = \begin{cases}
        rx  + (1-\zeta) w\varepsilon\ell(a) + b, \quad & \text{if } a < a_{\text{ret}},\\
        rx  + \varsigma(w) + b, & \text{if } a \geq a_{\text{ret}}.
    \end{cases}
\end{align}

At a fixed time $s$, consider a living household. Observing their current idiosyncratic state $(a, x, \varepsilon)$, they solve the following optimization problem
\begin{alignat}{2} \label{eq:olg_obj}
     V \coloneqq  \sup_{c}\; \E_s \Big[ {} & {} \int_{s}^{s+\tau}  e^{-\rho (t-s)} u(c_t) \,dt + e^{-\rho\tau}U(&&x_{s+\tau})  \Big]  \\ 
    \label{eq:olg_initcond_a}
    \text{s.t. }
    & a_s=a, &&
    \\
    \label{eq:olg_lawofmotion_a}
         & \Dot{a}_t = 1, \qquad && \forall t\in [s,s+\tau],
    \\
    \label{eq:olg_initcond_x}
         & x_s=x, &&
    \\
    \label{eq:olg_lawofmotion_x}
         & \Dot{x}_t = \iota(a_t, x_t, \varepsilon_t, \widetilde{r}_t, \widetilde{w}_t, \widetilde{b}_t) - c_t, \qquad && \forall t\in [s,s+\tau],
    \\
    \label{eq:olg_borrowingconstraint}
         & x_t \geq 0, \qquad && \forall t\in [s,s+\tau],
\end{alignat}
where $\mathbb{E}_s$ denotes the expectation conditional on the natural filtration generated jointly by the stochastic processes, and $\{\widetilde{r}_t\}$, $\{\widetilde{w}_t\}$, and $\{\widetilde{b}_t\}$ are household's belief processes for interest rate, wage, and received bequests, respectively. In a recursive rational expectations equilibrium, beliefs coincide with model reality and hence depend on the distribution of households, see below.

\paragraph{Distribution of households.}

The distribution of households over their idiosyncratic state variables is a key element in the model. Since the drift on age is positive and constant, it will be useful to distinguish two distribution objects -- the distribution over economic variables wealth and labor productivity
conditional on age, and the distribution on the entire idiosyncratic state space including age. In the remainder, let $A=[\underline{a}, \overline{a}]$ and $X=\mathbb{R}_{+}$.


First, within a generation of age $a$, there is a probability distribution over economic idiosyncratic state variables, $X\times \mathcal{E}$. We call the mapping from age to this distribution a \textit{generational distribution mapping} $\nu_t: A \rightarrow \mathcal{M}(X\times \mathcal{E})$. The distributions $\{\nu_t(a), a\in A\}$ are endogenously determined in equilibrium and describe how wealth and productivity are distributed within generations, conditional on their age.

Second, there is a measure $\mu_t$ over the entire idiosyncratic state space, $A\times X\times \mathcal{E}$. It determines the mass of households in the given idiosyncratic states. This measure satisfies, for each measurable function $\phi: A\times X\times \mathcal{E} \rightarrow \mathbb{R}$,
\begin{equation} \label{eq:distribution_mu}
    \int_{A\times X\times \mathcal{E}} \phi \,\mu_t(da,dx,d\varepsilon) = \int_A\int_{X\times \mathcal{E}}\phi\,\nu_t(a)(dx,d\varepsilon)  \lambda_t(da).
\end{equation}
Hence, the family $\{\nu_t(a), a\in A\}$ constitutes the disintegration of $\mu_t$ with respect to the demographic measure $\lambda_t$.

\paragraph{Bequest flows}
Equipped with notation of distributions defined above, we can state the formulas for bequest flows. The collected bequests can be written as Lebesgue density
\begin{align} \label{eq:model_b_out}
    b_t^{\text{out}}(a) & {} = \underbrace{\int_{X\times \mathcal{E}} (\1_{a<\overline{a}}\pi(a)+\delta_{\overline{a}}(a)) x \, \nu_t(a)(dx,d\varepsilon)}_{\text{collected bequests within generation $a$}} \cdot \underbrace{\vphantom{\int_{X\times Z}}f_t^{\lambda}(a),}_{\text{gen. size}}
\end{align}
where $\delta_{\overline{a}}$ denotes the Dirac delta at $\overline{a}.$ We distribute bequests as described above which results in the convolution formula
\begin{align} \label{eq:model_b_in}
    b_t^{\text{in}}(a) & {} = \frac{1}{f_t^{\lambda}(a)} \Big\{\underbrace{\int_0^{\overline{a}-a} b_t^{\text{out}}(a+t) p(t) \, dt}_{\text{bequest to children}} + \underbrace{b_t^{\text{out}}(a)\Big( 1 - \int_{0}^{a} p(t) dt\Big)}_{\text{bequest to siblings}}\Big\},
\end{align}
Note that no bequest is destroyed or created, $\int b^{\text{in}} f^\lambda da=\int b^{\text{out}} da$.\footnote{Conservation of bequest mass follows from a change of variables and Fubini's theorem.}

\paragraph{Firms.}

Production is determined by a representative firm with Cobb-Douglas production technology,
\begin{equation} \label{eq:CobbDouglas_production}
    F(K, L, Z) = \Gamma(Z) K^\alpha L^{1-\alpha} - \delta(Z) K,
\end{equation}
with the total factor productivity $\Gamma$ and depreciation rate $\delta$ depending on the aggregate shock $Z$. 

First-order conditions of the profit-maximizing firm require 
\begin{align}
    \label{eq:price_r}
    r_t = \partial_K F(K_t, L_t, Z_t), \\
    \label{eq:price_w}
    w_t =  \partial_L F(K_t, L_t, Z_t).
\end{align}

\paragraph{Equilibrium.}

In a recursive rational expectations equilibrium, households' choices are optimal given their beliefs over prices, households' beliefs are consistent with model reality, firms' choices are optimal given prices, and markets clear,
    \begin{align}
        K & {} =\int x \, d\mu(a,x,\varepsilon) \\
        L & {} = \int_{\underline{a}}^{a_{\text{ret}}}\int_{\mathcal{E}} \varepsilon \ell(a) \, \eta_{\varepsilon}(d\varepsilon) \, \lambda(da).
    \end{align}

\clearpage

\section{Solving OLG models with aggregate risk} \label{sec:SolvingAggRisk}
Consider the OLG model from section~\ref{sec:Model} with aggregate, but without idiosyncratic risk. That is, the set of idiosyncratic shock values is a singleton, $\mathcal{E}=\{\overline{\varepsilon}\}$, and the stationary distribution is a Dirac measure, $\eta_{\varepsilon} = \delta_{\overline{\varepsilon}}$. This setting retains the main difficulty -- aggregate risk in a rational expectations heterogeneous-agent model. At the same time, it simplifies the treatment of the infinite-dimensional distribution $\mu$ thanks to the following observation: without idiosyncratic risk, all agents within one generation are ex post identical, so we only need to keep track of the generation-wise asset holdings. In other words, the generational distribution $\nu(a)(dx,dz)$ concentrates on the graph of a function, the \textit{generational wealth function} $g: A \rightarrow X$. For all $a\in A$,
\begin{align} \label{eq:Solving_AggRisk_def_g}
    \nu(a)(dx,d\varepsilon) = (\delta_{g(a)}\otimes \delta_{\mathcal{E}})(dx, d\varepsilon).
\end{align}
Hence, from equation~\eqref{eq:Solving_AggRisk_def_g} together with ~\eqref{eq:distribution_mu}, the entire distribution $\mu\in\mathcal{M}(A\times X\times \mathcal{E})$ can be characterized by the function $g \in X^A \eqqcolon \mathcal{G}$.

\subsection{Equilibrium conditions}

We adopt recursive notion of equilibrium. Hence, optimal consumption $c^{\ast}$ and saving $s^{\ast}$, as well as value function $V$ depend on the recursive state $(a, x, z, g) \in A\times X \times \mathcal{Z} \times \mathcal{G}$. Since the set of aggregate shock realizations $\mathcal{Z}$ is finite, we write, e.g., $V_k(a,x,g) \coloneqq V(a,x,z_k, g)$, $k\in\{1, \dots, p\}$.

\paragraph{Master equation.} In equilibrium, $V$ satisfies the master equation
\begin{align} 
    \rho V_{k} &{} = \partial_a V_{k}  + H_k(\partial_x V_{k})  + \sum_{l=1}^{p} Q^{Z}_{kl} V_{l} + \pi(a)(U(x)-V_{k}) \nonumber \\
    &{} \hphantom{{}={}} + \int_A (\mathcal{T}_k g)(a')  \Big[\frac{\delta V_{k}}{\delta g}\Big](da'),
    \label{eq:AggRisk_MasterEquation}
\end{align}
with Hamiltonian $H_k(p| a,x, g) \coloneqq \sup_c\{u(c) + s_k(c|a,x,g)p\}$, and transport operator
\begin{align} \label{eq:AggRisk_TransportOperator}
    (\mathcal{T}_k g)(a) \coloneqq -g'(a) + s_k^{\ast}(a, g(a), g).
\end{align}
As in equation~\eqref{eq:olg_lawofmotion_x}, unoptimized savings are given by $s_k(c|a,x,g) \coloneqq \iota_k(a,x,g) - c$. We derive the above master equation formally in appendix~\ref{app:MasterEquation}.

\paragraph{Boundary conditions.} We impose boundary conditions
\begin{align}
    V_k(\overline{a},x,g) & {} = U(x), \label{eq:equilibrium_terminal_age}\\
    x & {} \geq 0. \label{eq:equilibrium_borrowingconstraint}
\end{align}
The terminal condition~\eqref{eq:equilibrium_terminal_age} is determined by the bequest motive while the state constraint~\eqref{eq:equilibrium_borrowingconstraint} corresponds to the borrowing constraint.

\subsection{Calibration} \label{subsec:SolvingAggRisk_Calibration}

Aggregate shocks to total factor productivity and depreciation are externally calibrated and chosen to deliver substantial aggregate volatility. All other parameters are either also externally calibrated or internally calibrated in the model without aggregate risk.\footnote{Equilibrium conditions for the model without aggregate risk are given by a system of (finite-dimensional) PDEs and corresponding boundary conditions and can be solved with classical techniques from \citet{achdou2022}. See appendix~\ref{app:SolveNoRisk} for details.} For the latter calibration, our strategy is analogous to \citet{denardi2004}.

\paragraph{Demographics and retirement system.} We consider the age interval $[\underline{a}, \overline{a}] = [20, 95]$. Death rates $\pi(a)$ are taken from the Human Mortality Database for the US in 2015 (HMD, {}\citeyear{HMD2025}). We set the distribution of parent ages $p(a)$ to a symmetric distribution between ages 20 and 40, see appendix~\ref{app:Calibration_external}. Retirement age is fixed at $a_{\text{ret}}=65$. The contribution is $\zeta=0.12$, roughly equal to the Social Security contribution rate in the US.

\paragraph{Households.}
We calibrate household parameters analogously to \citet{denardi2004}. Risk aversion is $\sigma=1.5$. The discount rate $\rho$ and the bequest utility parameters $\phi_1$ and $\phi_2$ are calibrated to match the following data: An interest rate of 3\%, a transfer wealth share\footnote{By transfer wealth share, we mean the share of aggregate wealth that is accumulated due to bequest flows, including interest. It is given by $(B/r)/K$, where $B=\int b^{\text{out}}(a)da$ is the aggregate bequest flow at each time which is constant without aggregate risk.} of 60\%, and a 30\% share of households with wealth of less than 6.25\% of the median income. The labor productivity profile is fitted to PSID data from 2015, see appendix~\ref{app:Calibration_external}.

\paragraph{Production and aggregate risk.}
As in \citet{denardi2004}, we choose $\alpha=0.36$. For the aggregate shocks $Z_t$, we set realizations of total factor productivity $\Gamma\in\{0.95, 1.05\}$ and depreciation $\delta\in\{0.08, 0.12\}$. In total, $Z_t$ has four realizations, consisting of the cross-product of both two-shock realizations. We choose jump rates such that jumps for $\Gamma$ and $\delta$ are independent and occur, in expectation, every ten years for $\Gamma$ and every three years for $\delta$, see table~\ref{tab:Calibration_AggRisk}.

\begin{table}[H]
    \footnotesize
    \centering
    \begin{adjustbox}{width=\textwidth}
    \begin{tabular}{ccll}
        \toprule
         \textbf{Parameter} & \textbf{Value} & \textbf{Description} & \textbf{Target / Source} \\
         \midrule
         \multicolumn{4}{l}{\textit{Demographics}} \\
         $[\underline{a}, \overline{a}]$ & [20, 95] & lifespan & ah-hoc \\
         $\pi(a)$ & see text & death rates & Human Mortality Database, US 2015.\\
         $p(a)$ & see text & parent-age distribution & ad-hoc, children are born between ages 20-40. \\[0.2cm] 
         
         \multicolumn{4}{l}{\textit{Household}} \\
         $\rho$ & 0.0241 & discount rate & interest rate $r\approx 3\%$ \\
         $l(a)$ & see text & labor profile & PSID, 2015. \\
         $\sigma$ & 1.5 & risk aversion & \citet{denardi2004} \\
         $\phi_1 $ & -7.4 & bequest desirability & \citet{denardi2004}, target: $(B/r)/K=0.6$\\
         $\phi_2 $ &  5.0 & bequest non-homo. & \citet{denardi2004}, target: $\text{cdf}_{\text{beq.}}(6.25\% \text{med. income}) = 30\%$ \\
         $\varepsilon$ & 1.0 & labor productivity (deterministic) & normalized \\
         $Q^{\varepsilon}$ & $[-0.0]$ & transition rates for $\varepsilon$ & no idiosyncratic risk \\[0.4cm]
        
         \multicolumn{4}{l}{\textit{Retirement system}} \\
         $a_{\text{ret}}$ & 65 & retirement age & ad-hoc \\
         $\zeta$ & 0.12 & contribution rate & ad-hoc \\[0.2cm]
         
         \multicolumn{4}{l}{\textit{Production}} \\
         $\alpha$ & 0.36 & capital share & \citet{denardi2004} \\ 
         $A(Z)$ & $[0.95, 0.95, 1.05, 1.05]$ & total factor productivity & ad-hoc \\
         $\delta(Z)$ & $[0.12, 0.08, 0.12, 0.08]$ & depreciation & ad-hoc \\
          $Q^{Z}$ & $\left[\begin{array}{cccc}
              - 0.43 & 0.33 & 0.1 & 0.0 \\
              0.33 & -0.43 & 0.1 & 0.0 \\
              0.1 & 0.0 & - 0.43 & 0.33 \\
              0.0 & 0.1 & 0.33 & -0.43 \\
         \end{array}\right]$ & transition rates for $Z$ & ad-hoc \\
         
         \bottomrule
    \end{tabular}
    \end{adjustbox}
    \caption{Parameters of the model with aggregate risk.}
    \label{tab:Calibration_AggRisk}
\end{table}

\subsection{Solution method} \label{subsec:SolvingAggRisk_SolutionMethod}
We use our general solution method from section~\ref{sec:GeneralSolutionMethod}. In the notation of section~\ref{sec:GeneralSolutionMethod}, we choose $X_1=A\times X \times \mathcal{Z}$ and $X_2= \mathcal{C}^0(A, X)$ for the low- and high-dimensional parts of the state space, respectively.\footnote{The model generates continuous $g$ functions in simulations as long as the initial distribution is continuous since one can bound the optimal savings away from $\pm\infty$.} Before applying our method, we need to approximate the infinite-dimensional distribution with a finite number of parameters.

\paragraph{Finite-dimensional distribution approximation.} Since the distribution $\mu\in\mathcal{M}(A,X)$ is characterized by the generational wealth function $g\in\mathcal{C}^{0}(A, X)$, we need to approximate only a continuous, one-dimensional function on a bounded domain. We do so on 
a space of linear splines with $n$ parameters 
and obtain $g_{\gamma} \in S^1_n(A,X)$,
with the vector of parameters $\gamma\in \mathbb{R}^n$. We choose a non-uniform discretization with $n=26$.\footnote{We perform first experiments with a uniform grid, and refine the grid in those regions where households leave or hit the borrowing constraint to capture the shape of $g$ more accurately.}

\paragraph{Finite-difference neural operator.}  

\begin{figure}
    \centering
    \includegraphics[width=1.0\textwidth]{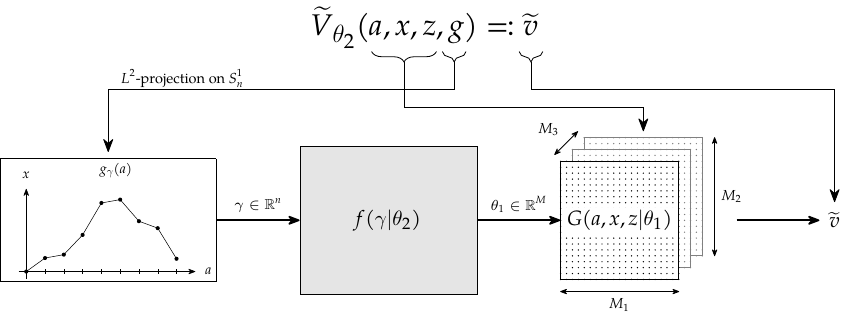}
    \caption{Finite-difference neural operator for the model with aggregate risk. This operator takes the projected generational wealth function $g_{\gamma}$ as an input and outputs a grid of values on the low-dimensional state space $A\times X \times \mathcal{Z}$. Finally, we obtain values at off-grid states by interpolation.}
    \label{fig:SolutionMethodAggRisk}
\end{figure}

We discretize the low-dimensional part of the state-space with a grid of sizes $M_1$, $M_2$, and $M_3$ in dimensions $A$, $X$, and $\mathcal{Z}$, respectively.\footnote{We choose the grid in the age dimension as refinement of the grid to approximate generational wealth functions. In particular, $M_1 \geq n$.} In this specific example, we truncate $X$ at a sufficiently large upper boundary $\overline{x}=30$ and set $M_1=151$ with uniform grid points, $M_2=155$ with decreasing mesh size towards the borrowing constraint, and $M_3=4$ as implied by the shock choice. Figure~\ref{fig:SolutionMethodAggRisk} sketches the resulting finite-difference neural operator. It takes an $n$-dimensional vector of parameters $\gamma$ as an input and outputs a vector of grid values of size $M=M_1M_2M_3$, conditional on $g_{\gamma}$. The resulting grid function $G\big(a,x,z\big| f_{\theta_2}(\gamma)\big)$ approximates the target value function $V(a,x,z,g_{\gamma})$.

\paragraph{Transport of distribution.}
During training, we use simulated data which requires us to transport any given $g_{\gamma}\in S^1_n(A,X)$. To capture the dynamics more accurately, we evaluate the discrete version of the transport operator~\eqref{eq:AggRisk_TransportOperator} on the (fine) age grid of size $M_1$ as chosen in the output grid. Consequently, however, the resulting, transported generational wealth function $g^{t+\Delta t}_{\gamma} \in S^1_{M_1}(A,X)$ does not need to be an element of $S^1_n(A,X)$. To approximate $g^{t+\Delta t}_{\gamma}$ in $S^1_{n}$, we choose the $L^2$-projection.

\paragraph{Neural net architecture.}
We implement the discrete operator from above $f_{\theta_2}$ as neural net in PyTorch \citep{ansel2024}. A thoughtful design of architecture (and training) is crucial for convergence of the overall method.
We use a standard feed-forward neural net. With the finite-dimensional distribution approximation and the finite-difference grid fixed (see above), our solution method prescribes input layer size $n$ and output layer size $M$. We are free to choose the number of hidden layers, their sizes, and the activation functions. We make a standard choice of three layers with size $128$, together with a CELU activation function \citep{barron2017}. To facilitate training, the neural net learns deviations from a reference solution -- a steady state value function $\overline{V}$ and corresponding generational wealth function $\overline{g}$. We compute this value function by fixing aggregate shock values to their mean levels and obtaining steady-state solution $\overline{V}$  to the corresponding heterogeneous-agent model without aggregate risk using the upwind finite-difference method from \citet{achdou2022}.

\paragraph{Shape preservation.} Training the neural net as is turned out to be fragile and delivered non-monotonic policy functions. This observation is in line with problems documented in \citet{gu2024}. To cure this pathological behavior, we propose a method to enforce concavity of the value function with respect to the wealth dimension. We do so by approximating second partial derivatives $\partial_{xx} V$ instead of plain values $V$.\footnote{To be precise, the neural net learns $\partial_{xx} V$ in the interior of $X$ plus two boundary values $V(\underline{x})$ and $V(\overline{x})$. Since partial derivatives and boundary values might be different in scales, we use two separate neural nets of the same size to approximate them.} Next, we enforce negativity of second derivatives using an exponential transformation. Let $y$ be the output given by the neural net. We define the approximated second partial derivative by $\partial_{xx} V \coloneqq \partial_{xx}\overline{V} e^{y}$, where $\overline{V}$ is the reference solution from above. Finally, backing out $V$ requires solving a tridiagonal system -- an operation that can be efficiently performed in $O(M_2)$. The resulting value function retains the sign in the second partial derivative of the reference solution; in our case, this implies concavity in $x$. In a similar way, one can enforce monotonicity of the neural net output. This is an important ingredient of the elegant method by \citet{azinovic-yang2026} which was developed independently of ours.

\paragraph{Training.} As in \citet{maliar2021} and \citet{azinovic2022}, we minimize squared residuals in equilibrium conditions using a stochastic gradient descent method. More precisely, we make the following standard choices. We initialize neural net weights with the Kaiming uniform method \citep{he2015} and update them iteratively using the ADAM optimizer \citep{kingma2017} with an exponential decay in the learning rate. We use a batch size of 32 and divide training in three phases. In the first phase, we construct artificial training data based on solutions to the deterministic problem.\footnote{We construct artificial data as convex combinations of solutions to the deterministic problem with the extreme realizations of the aggregate shock, with random weights.} After rough convergence, we enter the second phase and take a mixed batch of artificial and simulated data. Finally, once the loss has stabilized at low values, we switch to entirely simulated data. In all phases, we train three times on the same generated data, perturbed by a small noise to avoid overfitting.

\subsection{Results}

\begin{figure}
    \centering
    \begin{subfigure}{0.49\textwidth}
        \includegraphics[width=\textwidth]{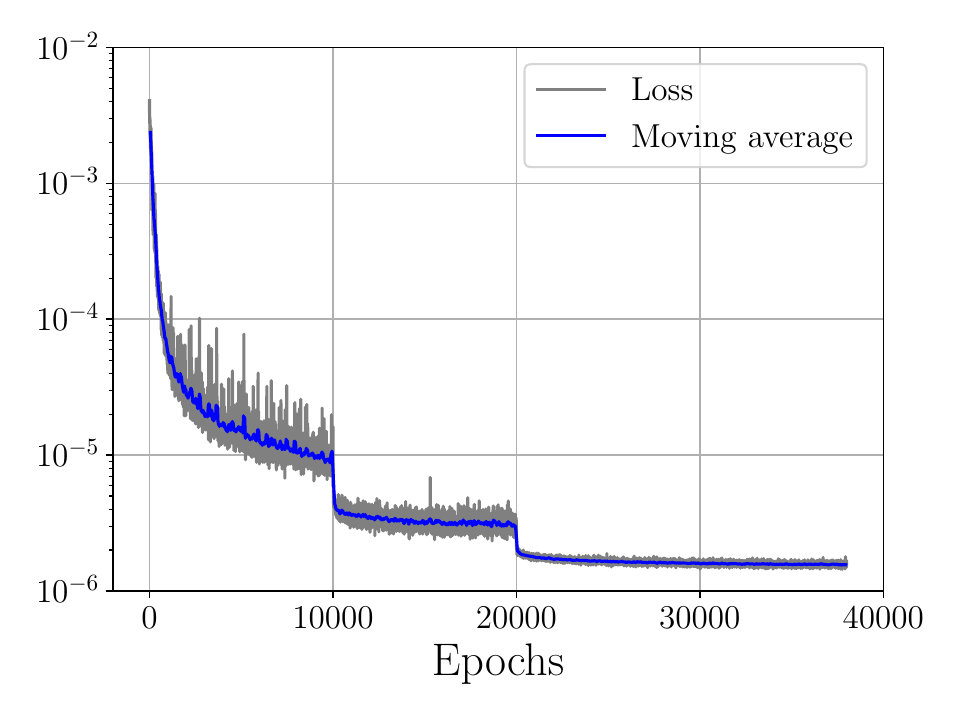}
        \label{fig:LossAndErgodicSet_AggRisk_loss} 
    \end{subfigure}
    \begin{subfigure}{0.49\textwidth}
        \includegraphics[width=\textwidth]{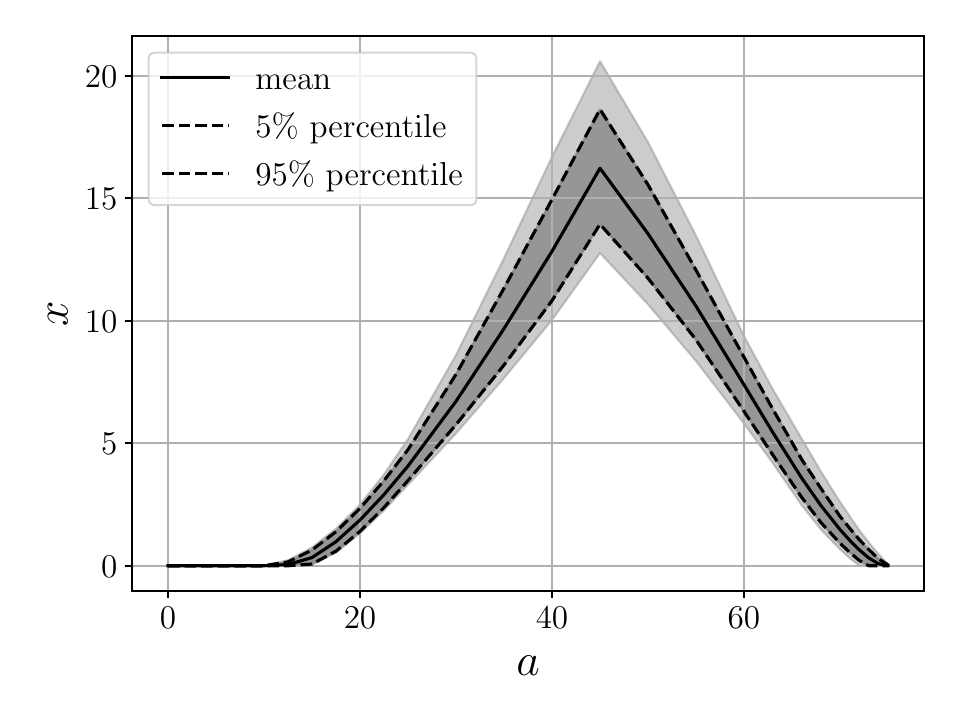}
        \label{fig:LossAndErgodicSet_AggRisk_ergodic_set} 
    \end{subfigure}
    \caption{Training loss (left) and approximated ergodic set (right) for the OLG model with aggregate risk. The three marked loss regimes reflect the fact that, during training, we go from a fully artificial data set with large noise magnitude to fully simulated data with minimal noise. The approximated ergodic set is visualized using minimum and maximum wealth values together with mean and 5th/95th-percentiles, pointwise in age. This confirms that the resulting value function implies stable but substantial dynamics in the generational wealth function. Saving occurs relatively late in the working life due to the labor profile and the timing of bequests reception; the borrowing constraint binds for the youngest generations. The oldest generations dissave their entire wealth, suggesting that the calibrated bequest motive is dominated by consumption utility in all shock realizations.}
    \label{fig:LossAndErgodicSet_AggRisk}
\end{figure}

Figure~\ref{fig:LossAndErgodicSet_AggRisk} summarizes the results of our finite-difference neural operator. Overall training takes about eight hours on a high-performance Nvidia H200 GPU. Panel (a) confirms that the loss converges to values of the order of $10^{-6}$.
Moreover, the resulting value function implies stable dynamics, as suggested by Panel (b): the set of simulated generational wealth functions (i.e., distributions) shows substantial variation, but remains bounded within reasonable wealth ranges. 

To assess accuracy, we construct a set of residuals observed over a set of simulated generational wealth functions and all points from the finite-difference grid. Mean and 99.9\% percentile of these residuals remain at low values of $0.0008$ and $0.0045$, respectively. Importantly, we do not observe systematic errors around the borrowing constraint and saving policies satisfy $s^{\ast}(x=0)\geq0$ up to discretization error, as confirmed by spatial error plots in appendix~\ref{app:AddResults_AggRisk}. This suggests that our approximation is able to capture the borrowing constraint well, which is a considerable challenge \citep{gu2024}.

\clearpage

\section{Solving OLG models with aggregate and idiosyncratic risk} \label{sec:SolvingAggIdioRisk}

In this section, we turn to the full model with aggregate and idiosyncratic risk. In this case, we cannot simplify the distribution as in section~\ref{sec:SolvingAggRisk}. Hence, equilibrium conditions will be given in terms of the full distribution $\mu$. This will pose additional challenges on the finite-dimensional approximation of the distribution -- however, once discretized, our finite-difference neural operator is constructed in the same way as in the previous case. We therefore keep this section focussed on the novel ingredients.

\subsection{Equilibrium conditions}

Optimal consumption $c^{\ast}$ and saving $s^{\ast}$, as well as the value function $V$ depend on the recursive state $(a, x, \varepsilon, z, \mu) \in A\times X \times \mathcal{E}  \times \mathcal{Z} \times \mathcal{M}$. As before, stochastic processes $(\varepsilon_t)$ and $(Z_t)$ have a finite number of realizations, and we write, $V_{ik}(a,x, \mu) =V(a,x,\varepsilon_i, z_k, \mu)$, and analogously for other variables. We omit dependencies when there is no risk of ambiguity.

\paragraph{Master equation.} Equilibrium is characterized by the master equation (cf. Appendix~\ref{app:MasterEquation})
\begin{align} 
    \rho V_{ik} &{} = \partial_a V_{ik}  + H_{ik}(\partial_x V_{ik}) + \sum_{j=1}^{q} Q^{\varepsilon}_{ij} V_{jk} + \sum_{l=1}^{p} Q^Z_{kl} V_{il} +\pi(a)(U(x)-V_{ik}) \nonumber \\
    &{} \hphantom{{}={}} + \int_A \int_X \sum_{j=1}^{q}\frac{\delta V_{ik}}{\delta \mu_j}(a', x') \big[\mathcal{T}_k \mu_j\big](da', dx'),
    \label{eq:AggIdioRisk_MasterEquation}
\end{align}
with Hamiltonian $H_{ik}(p| a,x,\mu) \coloneqq \sup_c\{u(c) + s_{ik}(c|a,x,\mu)p\}$, and transport operator
\begin{align} \label{eq:AggIdioRisk_TransportOperator}
     \mathcal{T}_k \mu_i & {} \coloneqq -\partial_a \mu_i -\partial_x[\mu_i s^{\ast}_{ik}] + \sum_{j=1}^{q} Q_{ji} \mu_j - \pi(a)\mu_i.
\end{align}
As in equation~\eqref{eq:olg_lawofmotion_x}, unoptimized savings $s$ are given by $s_{ik}(c|a,x,\mu) \coloneqq \iota_{ik}(a,x,\mu) - c$.

\paragraph{Boundary conditions.}
We impose boundary conditions
\begin{align}
    V_{ik}(\overline{a},x,\mu) & {} = U(x), \label{eq:AggIdioRisk_equilibrium_terminal_age}\\
    x & {} \geq 0. \label{eq:AggIdioRisk_equilibrium_borrowingconstraint}
\end{align}
The terminal condition~\eqref{eq:AggIdioRisk_equilibrium_terminal_age} is determined by the bequest motive while the state constraint~\eqref{eq:AggIdioRisk_equilibrium_borrowingconstraint} corresponds to the borrowing constraint.

\subsection{Calibration} \label{subsec:SolvingAggIdioRisk_Calibration}

Our calibration is analogous to section~\ref{subsec:SolvingAggRisk_Calibration}. 
Table~\ref{tab:Calibration_AggIdioRisk} lists the calibrated parameter values.

\begin{table}
    \footnotesize
    \centering
    \begin{adjustbox}{width=\textwidth}
    \begin{tabular}{ccll}
        \toprule
         \textbf{Parameter} & \textbf{Value} & \textbf{Description} & \textbf{Target / Source} \\
         \midrule
         \multicolumn{4}{l}{\textit{Demographics}} \\
         $[\underline{a}, \overline{a}]$ & [20, 95] & lifespan & ah-hoc \\
         $\pi(a)$ & see text & death rates & Human Mortality Database, US 2015.\\
         
         \multicolumn{4}{l}{\textit{Household}} \\
         $\rho$ & 0.068 & discount rate & interest rate $r\approx 3\%$ \\
         $l(a)$ & see text & labor profile & PSID, 2015 \\
         $p(a)$ & see text & parent-age profile & ad-hoc, children are born between biological ages 20-40. \\
         $\sigma$ & 1.5 & risk aversion & \citet{denardi2004} \\
         $\phi_1 $ & -26.0 & bequest desirability & \citet{denardi2004}, target: $(B/r)/K=0.6$\\
         $\phi_2 $ &  6.1 & bequest non-homo. & \citet{denardi2004}, target: $\text{cdf}_{\text{beq.}}(6.25\% \text{med. income}) = 30\%$ \\
         $\varepsilon$ & $[0.354, 2.829 ]^{\top}$ & labor productivity &\citet{denardi2004}, transformed 5-year AR(1) \\&&& \quad to Ornstein-Uhlenbeck process with discretized innovations \\
         $Q^{\varepsilon}$ & $\left[\begin{array}{cc}
              - 0.033 & 0.033  \\
              0.033 & -0.033
         \end{array}\right]$ & transition rates for $\varepsilon$ & \citet{denardi2004}, transformed as above \\[0.4cm]
        
         \multicolumn{4}{l}{\textit{Retirement system}} \\
         $a_{\text{ret}}$ & 65 & retirement age & ad-hoc \\
         $\zeta$ & 0.12 & contribution rate & ad-hoc \\[0.2cm]

         \multicolumn{4}{l}{\textit{Production}} \\
         
         $\alpha$ & 0.36 & capital share & \citet{denardi2004} \\ 
         $A(Z)$ & $[0.95, 0.95, 1.05, 1.05]$ & total factor productivity & ad-hoc \\
         $\delta(Z)$ & $[0.12, 0.08, 0.12, 0.08]$ & depreciation & ad-hoc \\
          $Q^{Z}$ & $\left[\begin{array}{cccc}
              - 0.43 & 0.33 & 0.1 & 0.0 \\
              0.33 & -0.43 & 0.1 & 0.0 \\
              0.1 & 0.0 & - 0.43 & 0.33 \\
              0.0 & 0.1 & 0.33 & -0.43 \\
         \end{array}\right]$ & transition rates for $Z$ & ad-hoc \\
         
         \bottomrule
    \end{tabular}
    \end{adjustbox}
    \caption{Calibration of the model with idiosyncratic risk. Our calibration strategy is, in large parts, analogous to \citet{denardi2004}.}
    \label{tab:Calibration_AggIdioRisk}
\end{table}

\subsection{Solution method}
\label{subsec:SolvingAggIdioRisk_SolutionMethod}

This section parallels the descriptions for the case with aggregate risk alone from section~\ref{subsec:SolvingAggRisk_SolutionMethod}. The main complication is, as mentioned, the approximation of distribution $\mu$.

In the notation of section~\ref{sec:GeneralSolutionMethod}, we choose $X_1=A\times X \times \mathcal{E} \times \mathcal{Z}$ and $X_2 = \mathcal{M}(A\times X\times \mathcal{E})$ for the low- and high-dimensional parts of the state space, respectively. Again, we need to approximate $X_2$ in finite dimensions.

\paragraph{Finite-dimensional distribution approximation.}

We approximate $\mu$ with a density $m$ on the same grid as used for the finite-difference neural operator below. Hence $m$ has $M_1M_2M_3$ parameters. Since this number is, in practice, substantially larger than what is feasible for our method, we propose a further dimensionality reduction, to $m_{\gamma} \approx m$, with $\gamma \in \mathbb{R}^n$.\footnote{The number of input parameters $n$ must not be too high for the reasons outlined in footnote~\ref{fn:GeneralMethod_inputs_not_too_large}. In the subsequent computation, $n \leq 200$, while $M_1M_2M_3=61,910$.} A detailed description of our dimensionality reduction technique is postponed to appendix~\ref{app:DistributionApprox}.

\paragraph{Finite-difference neural operator.}  

\begin{figure}
    \centering
    \includegraphics[width=1.0\textwidth]{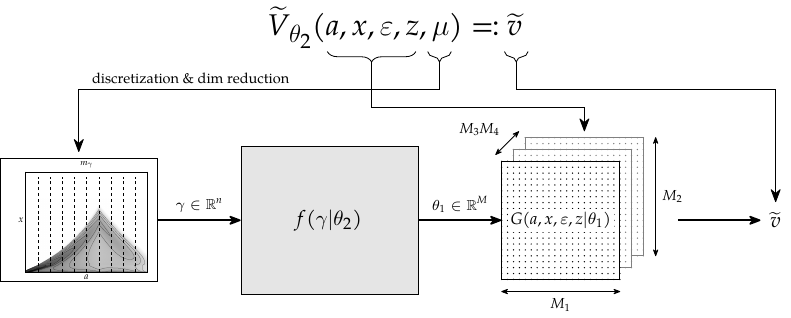}
    \caption{Finite-difference neural operator for the model with aggregate risk. This operator takes the distribution approximation $m_{\gamma}$ as an input and outputs a grid of values on the low-dimensional state space $A\times X \times \mathcal{Z}$. Finally, we obtain values at off-grid states ($a,x,\varepsilon, z$) from grid values $\theta_1$ by interpolation.}
    \label{fig:SolutionMethodAggIdioRisk}
\end{figure}

We discretize the low-dimensional part of the state-space with a non-uniform grid of sizes $M_1$, $M_2$, $M_3$, and $M_4$ in dimensions $A$, $X$, $\mathcal{E}$, and $\mathcal{Z}$, respectively. We truncate $X$ at a sufficiently large upper limit $\overline{x}=55$ and set $M_1=151$ with uniform grid points, $M_2=205$ with decreasing mesh size towards the borrowing constraint, as well as $M_3=2$ and $M_4=4$ as implied by the shock choices. The upwind finite difference scheme and in particular the discrete version of the transport operator from equation~\eqref{eq:AggIdioRisk_TransportOperator} are constructed taking into account the non-uniformity of the grid in the $x$ variable. Figure~\ref{fig:SolutionMethodAggIdioRisk} sketches the resulting finite-difference neural operator. It takes an $n$-dimensional vector of parameters $\gamma$ as an input and outputs a vector of grid values of size $M=M_1M_2M_3M_4$, conditional on $m_{\gamma}$. The resulting grid function $G\big(a,x,\varepsilon, z\big| f(\gamma|\theta_2)\big)$ approximates the target value function $V(a,x,\varepsilon, z,m_{\gamma})$.

\paragraph{Training.} Since the law of motion implied by the initial value function turned out to be rather well-behaved, we directly start training from simulated data.\footnote{So far, we suspect that the simulation with both risks is better behaved due to the simulated path being \enquote{smoother} than in the case with the degenerate measure.} That is, we immediately start with training phase 3 from above.

\subsection{Results}
\begin{figure}
    \centering
    \begin{subfigure}{0.49\textwidth}
        \includegraphics[width=\textwidth]{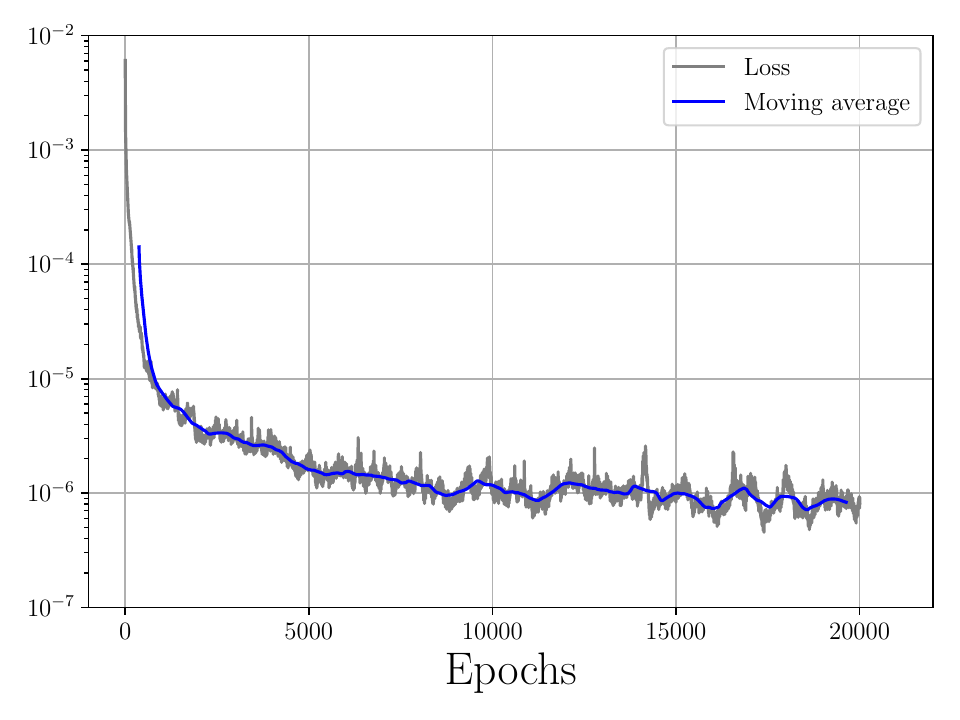}
        \label{fig:LossAndErgodicSet_AggIdioRisk_loss} 
    \end{subfigure}
    \begin{subfigure}{0.49\textwidth}
        \includegraphics[width=\textwidth]{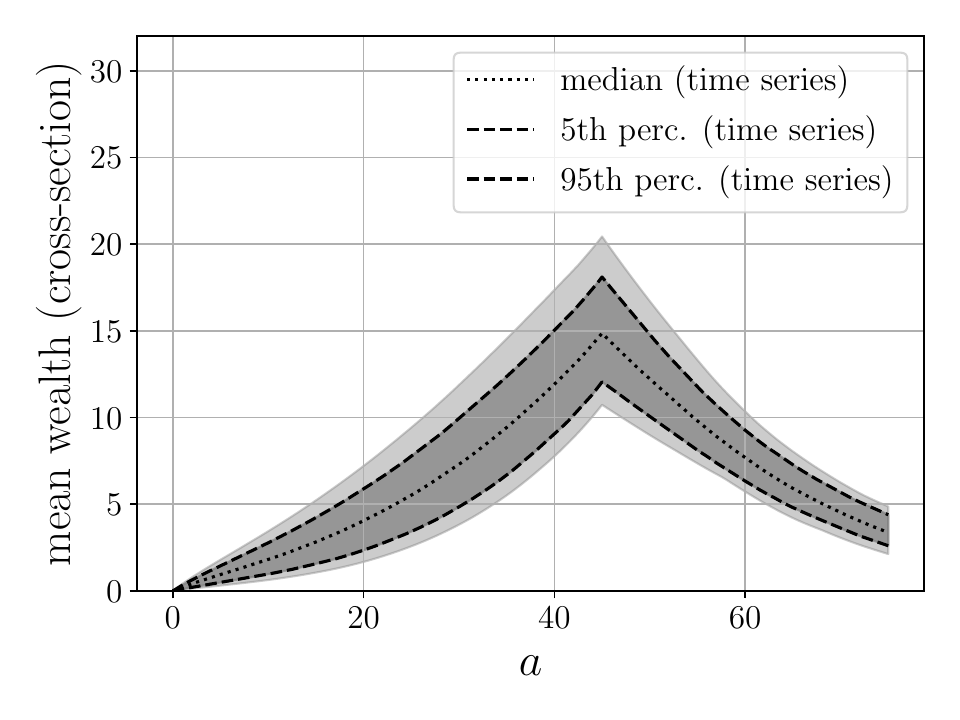}
        \label{fig:LossAndErgodicSet_AggIdioRisk_ergodic_set} 
    \end{subfigure}
    \caption{Training loss (left) and approximate ergodic distribution of age-specific mean wealth (right) for the OLG model with aggregate and idiosyncratic risk. More precisely, the right plot displays time-series percentiles, as well as min and max, over long simulations. The graph confirms that the computed value function implies substantial dynamics in the wealth distribution. Compared to the case without idiosyncratic risk, households hold, on average, more wealth when young because of the additional precautionary savings motive.}
    \label{fig:LossAndErgodicSet_AggIdioRisk}
\end{figure}

\begin{figure}
    \centering
    \begin{subfigure}{0.49\textwidth}
        \includegraphics[width=\textwidth]{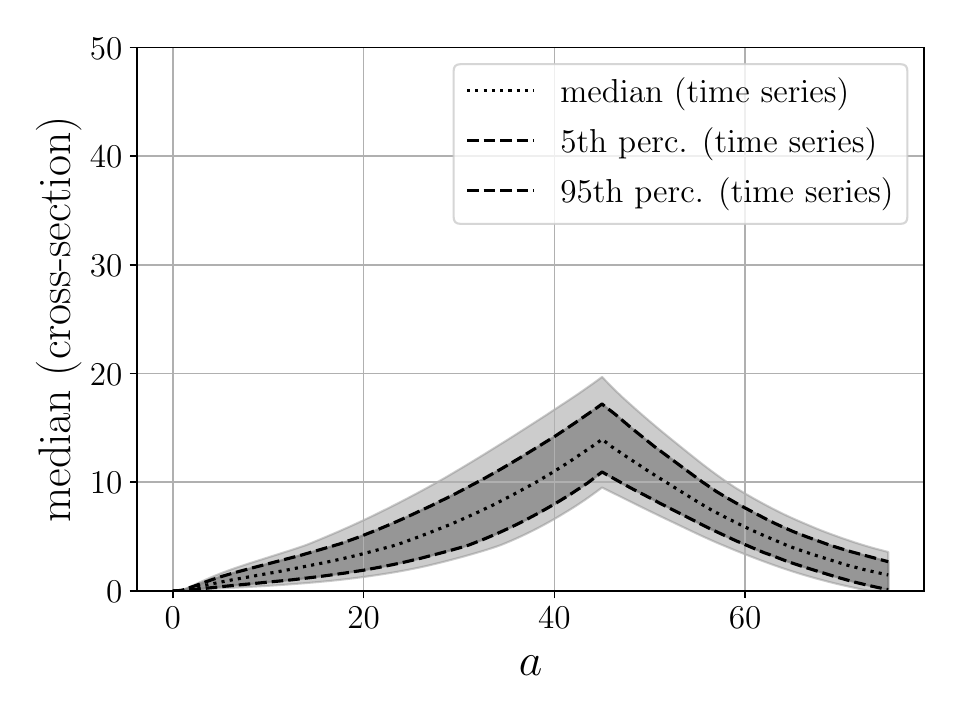}
        \caption{}
        \label{fig:LossAndErgodicSet_AggIdioRisk_ergodic_set_perc50} 
    \end{subfigure}
    \begin{subfigure}{0.49\textwidth}
        \includegraphics[width=\textwidth]{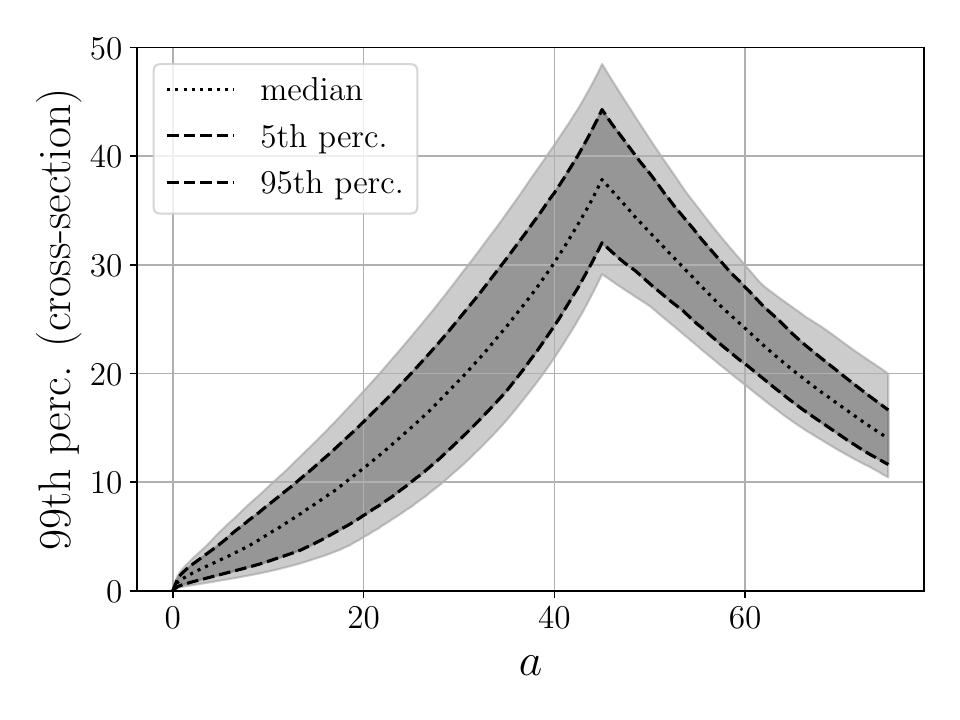}
        \caption{}
        \label{fig:LossAndErgodicSet_AggIdioRisk_ergodic_set_perc99} 
    \end{subfigure}
    \caption{Variation in the cross-sectional wealth median (a) and 99th percentile (b) of the approximated ergodic set in the OLG model with aggregate and idiosyncratic risk. More precisely, the plots display time-series percentiles of said quantities, as well as min and max, over long simulations.
    }
    \label{fig:VariationInErgodicSet_AggIdioRisk}
\end{figure}

Figure~\ref{fig:LossAndErgodicSet_AggIdioRisk} shows the results of our finite-difference neural operator, after 22 hours of training on a high-performance Nvidia H200 GPU. As in the case without idiosyncratic risk, the loss converges to values of the order of $10^{-6}$, see panel (a). Panel (b) suggests substantial but stable dynamics-; the cross-sectional mean wealth, conditional on age, shows substantial variation, but remains bounded within reasonable wealth ranges. This finding is confirmed when looking at 50th and 99th percentiles of wealth, see figure~\ref{fig:VariationInErgodicSet_AggIdioRisk}.\footnote{We checked other statistics to verify stability of the simulations, too. For instance, aggregate capital realizes within a range of approximately $[3,8]$.}

To assess accuracy, we construct a set of residuals observed over a set of simulated approximate densities $m_{\gamma}$ and all points from the finite-difference grid. Mean and 99.9\% percentile of these residuals remain at low values of $0.0005$ and $0.0068$, respectively.\footnote{This error measure is a mix of simulation and state space error. It is the natural way to assess accuracy since our method is defined on simulated paths in the distribution and on the full state space in low-dimensional variables.} As in the case without aggregate risk, our results suggest that the borrowing constraint is well-captured by our method, see appendix~\ref{app:AddResults_AggIdioRisk}.


\clearpage
\printbibliography

\clearpage

\appendix

\section{Solving the model without risk}\label{app:SolveNoRisk}

In this section, we consider the continuous-time OLG model from section~\ref{sec:Model} without idiosyncratic and aggregate risk, i.e., its deterministic version. This model will serve two purposes; first for internal calibration in section~\ref{subsec:SolvingAggRisk_Calibration} and second as a reference solution $\overline{V}$ in the solution method from section~\ref{subsec:SolvingAggRisk_SolutionMethod}.

\subsection{Equilibrium conditions}
We use the same notation as in the main part. Equilibrium for the deterministic model is characterized by the forward-backward system of an HJB equation and an ODE on the generational wealth function\footnote{This ODE is implied by the KFP equation that one typically expects in these applications.}
\begin{align}
    \label{eq:AggRisk_HJB} \rho V &{} = \partial_a V  + H(\partial_x V) +\pi(a)(U(x)-V), \\
    \label{eq:AggRisk_KF} 0 &{}= -g'(a) + s^{\ast}(a, g(a)), \\
    \label{eq:AggRisk_HJB_InitCond} V(\overline{a}, x) &{}= U(x), \\
    \label{eq:AggRisk_KF_InitCond} g(\underline{a}) &{} = 0.
\end{align}

\subsection{Solving the PDE system}
We solve the system \eqref{eq:AggRisk_HJB}-\eqref{eq:AggRisk_KF_InitCond} using upwind finite-differences from~\citet{achdou2022}.

\textit{-- TO BE COMPLETED --}

\clearpage

\section{Solving the model with idiosyncratic risk} \label{app:SolveIdioRisk}

In this section, we consider the continuous-time OLG model from section~\ref{sec:Model} with idiosyncratic risk alone, i.e., without aggregate risk. This model will serve two purposes; first for internal calibration in section~\ref{subsec:SolvingAggIdioRisk_Calibration} and second as a reference solution $\overline{V}$ in the solution method from section~\ref{subsec:SolvingAggIdioRisk_SolutionMethod}.

\subsection{Equilibrium conditions}
We use the same notation as in the main part. Equilibrium for the model with idiosyncratic risk only is characterized by the forward-backward system of HJB and KFP equation \citep{achdou2022}
\begin{align}
    \label{eq:AggIdioRisk_HJB} \rho V_i &{} = \partial_a V_i  + H(\partial_x V_i) + \sum_{j=1}^{q} Q ^{\varepsilon}_{ij} V_j +\pi(a)(U(x)-V_i), \\
    \label{eq:AggIdioRisk_KF} 0 &{}= -\partial_a \mu_i -\partial_x[\mu_i s_i] + \sum_{j=1}^{q} Q^{\varepsilon}_{ji} \mu_j - \pi(a)\mu_i, \\
    \label{eq:AggIdioRisk_HJB_InitCond} V_i(\overline{a}, x) &{}= U(x), \\
    \label{eq:AggIdioRisk_KF_InitCond} \nu(\underline{a}) &{} = (\delta_0 \otimes \eta_{\varepsilon})(dx, d\varepsilon), \quad \text{and} \quad f^{\lambda}(\underline{a}) = \frac{1}{\overline{a}-\underline{a}}.
\end{align}
Note that, somewhat unusually, we formulate the initial condition for $\mu$ in terms of $\nu(a)$ together with the generational density $f_{\lambda}$ since we want to prescribe the measure $\mu$ conditional on $a=\underline{a}$.

\subsection{Solving the PDE system}
We solve the system \eqref{eq:AggIdioRisk_HJB}-\eqref{eq:AggIdioRisk_KF_InitCond} using upwind finite-differences from~\citet{achdou2022}.

\textit{-- TO BE COMPLETED --}

\clearpage

\section{Calibration details} \label{app:Calibration}

This section details on the external calibration for all model versions as well as the internal calibration in the model without risk and in the model with idiosyncratic risk alone.

\subsection{External calibration}  \label{app:Calibration_external}

\begin{figure}[H]
    \centering
    \begin{subfigure}{0.32\textwidth}
    \includegraphics[width=\linewidth]{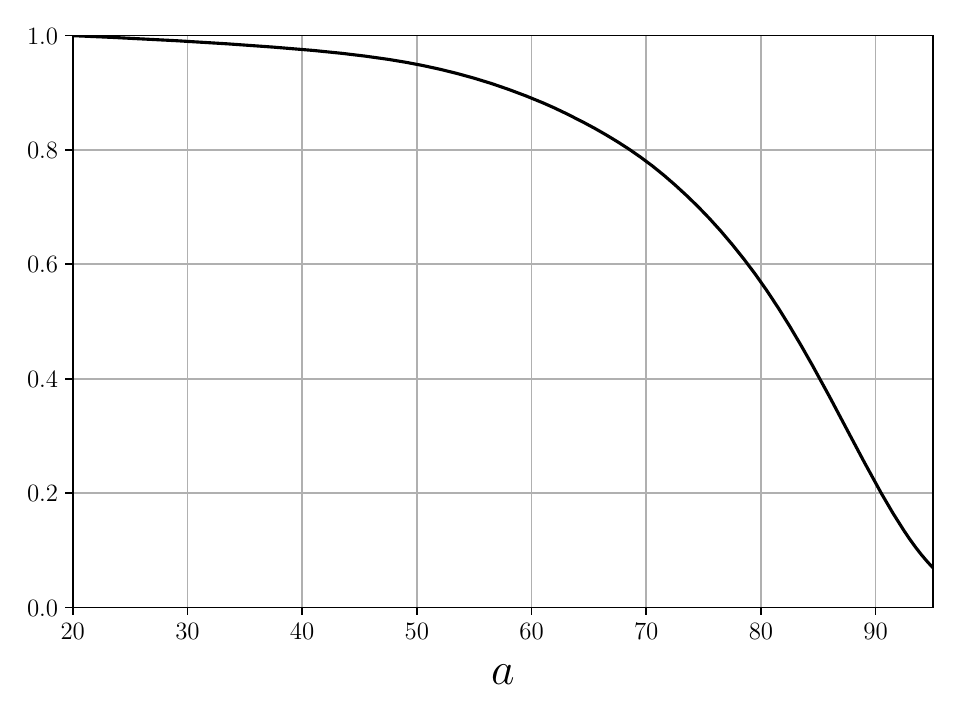}
    \subcaption{Demographic density $f^{\lambda}(a)$}
    \label{fig:ParameterProfiles_AggRisk_Demographics}
    \end{subfigure}
    \begin{subfigure}{0.32\textwidth}
    \includegraphics[width=\linewidth]{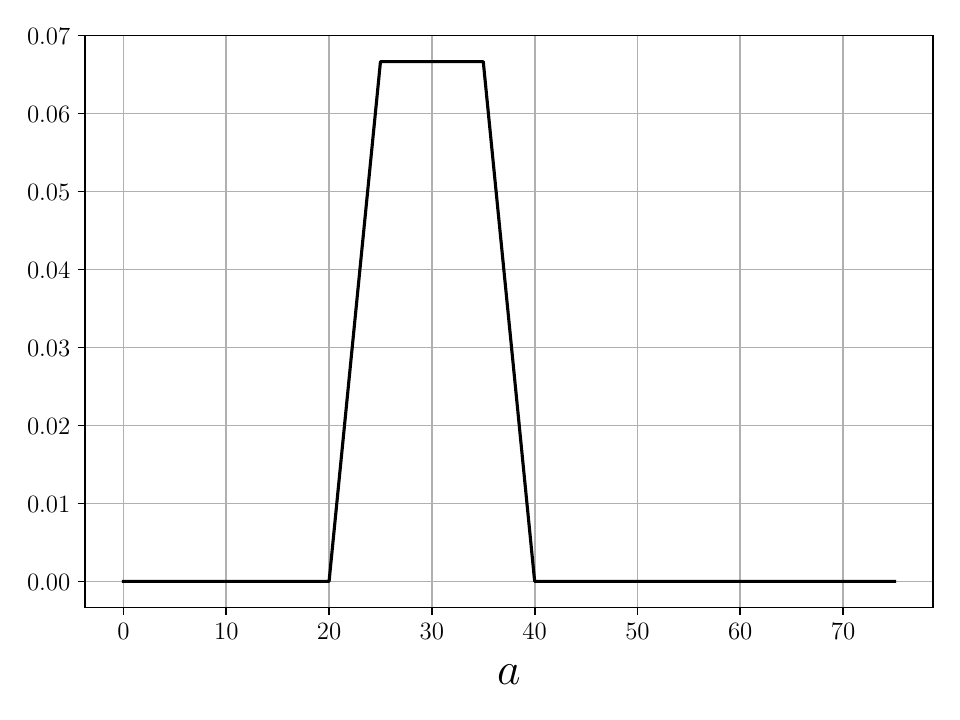}
    \subcaption{Parent-age density $p(a)$}
    \label{fig:ParameterProfiles_AggRisk_ParentAge}
    \end{subfigure}
    \begin{subfigure}{0.32\textwidth}
    \includegraphics[width=\linewidth]{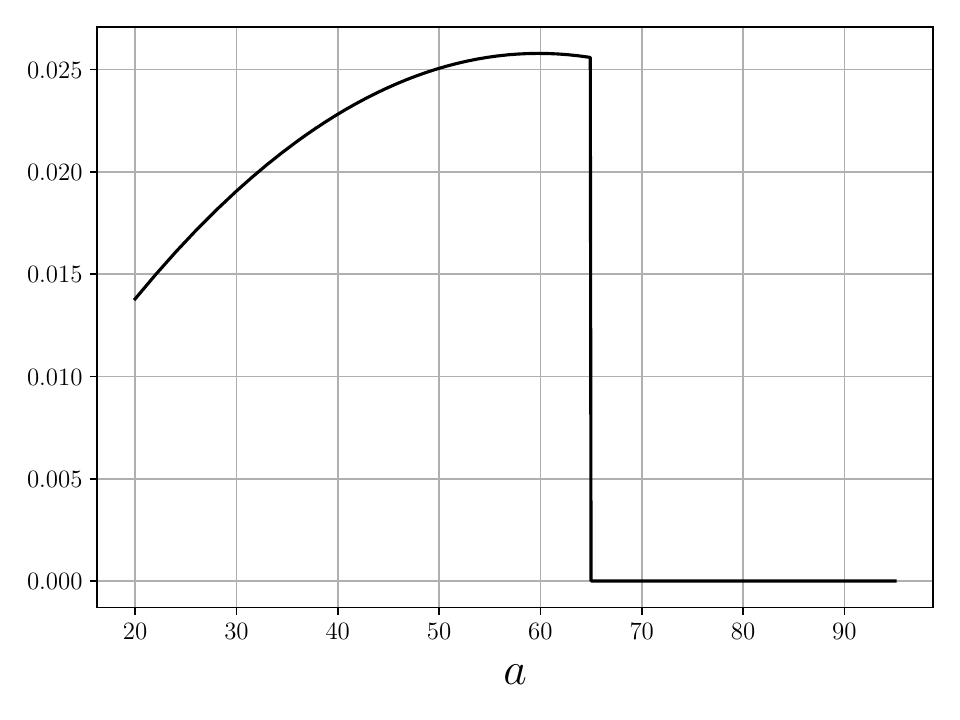}
    \subcaption{Labor profile $\ell(a)$}
    \label{fig:ParameterProfiles_AggRisk_LaborProfile}
    \end{subfigure}
    \caption{Choice of age-dependent parameter profiles. Parent-age density is normalized such that it integrates to one. The labor profile is normalized such that aggregate labor supply integrates to one. All plots are with respect to biological age $a$.}
    \label{fig:ParameterProfiles_AggRisk}
\end{figure}

\paragraph{Death probabilities.}
We take death probabilities from the Human Mortality Database for the US in 2015 (HMD, \citeyear{HMD2025}). We interpolate death rates linearly between integer ages. The resulting demographic density is shown in figure~\ref{fig:ParameterProfiles_AggRisk_Demographics}.

\paragraph{Parent-age distribution.} 
We choose a symmetric density between ages 20 and 40 to model the inter-generational links. Our simple approach uses a constant function between ages 25 and 35 and linearly interpolates down to zero on the remaining parts. In principle, one could fit a functional form to real data. However, since results turned out to be insensitive to reasonable choices of $p(a)$, we stick with the simple approach. The corresponding density is depicted in figure~\ref{fig:ParameterProfiles_AggRisk_ParentAge}.

\paragraph{Labor profile.}
We fit the labor productivity profile to 2015 data from the Panel Study of Income Dynamics (PSID, \citeyear{PSID2026}). We take labor productivity as the hourly regular wage rate as defined in the PSID and fit a quadratic polynomial. We include ages between 20 and 65.\footnote{The actual procedure used so far deviates slightly from the one described above and will be updated in the next version of this working paper. Differences in the implied labor profiles are quantitatively negligible.} The resulting normalized productivity profile is shown in figure~\ref{fig:ParameterProfiles_AggRisk_LaborProfile}.

\subsection{Internal calibration of the model without risk} \label{app:Calibration_InternalNoRisk}

\textit{-- TO BE COMPLETED --}

\subsection{Internal calibration of the model with idiosyncratic risk}  \label{app:Calibration_InternalIdioRisk}

The model with idiosyncratic risk only can be solved via a finite-difference upwind method for the system \eqref{eq:AggIdioRisk_HJB}-\eqref{eq:AggIdioRisk_KF_InitCond}, as described in appendix~\ref{app:SolveIdioRisk}. Calibration delivers parameter values from table~\ref{tab:Calibration_AggIdioRisk}. The calibrated model does a fairly good job in matching the empirical age-wealth shares, as table~\ref{tab:MomentsAgeWealthShares_AggIdioRisk} suggests.

\begingroup
\renewcommand{\arraystretch}{1.2}
\begin{table}[H]
    \centering
    \begin{tabular}{l r r}
        \toprule
        \multirow{2}{*}{Age group} & \multicolumn{2}{c}{ Wealth share} \\
        \cline{2-3} & model & data \\
         \midrule
         below 40 & 0.091 & 0.051 \\
         40-54 & 0.236 & 0.234 \\
         55-69 & 0.422 & 0.472\\
         above 70 & 0.251 &  0.242 \\
         \bottomrule
    \end{tabular}
    \caption{Untargeted age-wealth shares in the model and in the data. We report 2015 data from the FED. Age is given as biological age.}
    \label{tab:MomentsAgeWealthShares_AggIdioRisk}
\end{table}
\endgroup

\textit{-- TO BE COMPLETED --}

\clearpage

\section{Formal derivation of the Master equation}\label{app:MasterEquation}

In this section, we formally derive the master equation for the case with aggregate risk alone. To do so, we start from the case with both risks, where the derivation is close to \citet{cardaliaguet2019}. We then specialize the master equation to the case where the distribution $\mu \in \mathcal{M}(A\times X \times \mathcal{E})$ can be characterized via a generational wealth function $g\in\mathcal{G}=X^A$.

\textit{-- TO BE COMPLETED --}

\clearpage

\section{Distribution approximation}\label{app:DistributionApprox}

In this section, we describe our dimensionality reduction strategy for the discrete density $m$. This density is defined on a cartesian-product grid over $A\times X \times \mathcal{E}$ of size $M_1M_2M_3$. Since $\mathcal{E}$ is finite with $p$ values, we treat the approximation of $m_i=m(a,x,\varepsilon_i)$ independent of each other. Moreover, because the drift in age is constant, it is natural to approximate $m$ in age slices.

\paragraph{Approximation along age.}
We exploit the fact that policies are, in large parts of the domain, smooth in age.\footnote{In our model, the sole exception is retirement, which causes a drift discontinuity at retirement age.} We therefore approximate the conditional densities $m_{i}(a_l, \cdot)$ at a small set of ages $\{a_l\}_{l=1}^L$ and interpolate said conditional densities between these ages. Once the conditional distributions are approximated at all age slices (see below), we interpolate to arbitrary ages by \enquote{displacement interpolation}, a method from optimal transport that interpolates distributions along the shortest path in Wasserstein space \citep{gangbo1996, peyre2020}.\footnote{This approach is relatively cheap in one dimension. In higher dimensions, however, computing the optimal transport map is a challenging task.} We perform said interpolation on normalized densities and rescale with exogenous generation masses afterwards.

\paragraph{Approximation along wealth.} 
We perform the approximation of the one-dimensional $m_{i}(a_l, \cdot)$ in two steps. A single parameter $\gamma_0$ describes the mass contained in the interval $[0,\epsilon]$ (near the lower bound), and the part of the measure supported in $(\epsilon, \bar x]$ is projected on the space of exponentiated polynomials of degree $n-2$, yielding parameters $\gamma_1, \dots, \gamma_{n-1}$. This ensures that the resulting density is non-negative; and a further normalization guarantees mass preservation.

In total, our method approximates $m$ with $Lnp$, parameters, where $L < M_1$, $n < M_2$, and $p=M_3$.

\clearpage

\section{Additional results} \label{app:AddResults}

\subsection{Errors in the case with aggregate risk}\label{app:AddResults_AggRisk}

\textit{-- TO BE COMPLETED --}

\subsection{Errors in the case with aggregate and idiosyncratic risk}\label{app:AddResults_AggIdioRisk}

\begin{figure}[H]
    \centering
    \includegraphics[width=0.8\linewidth]{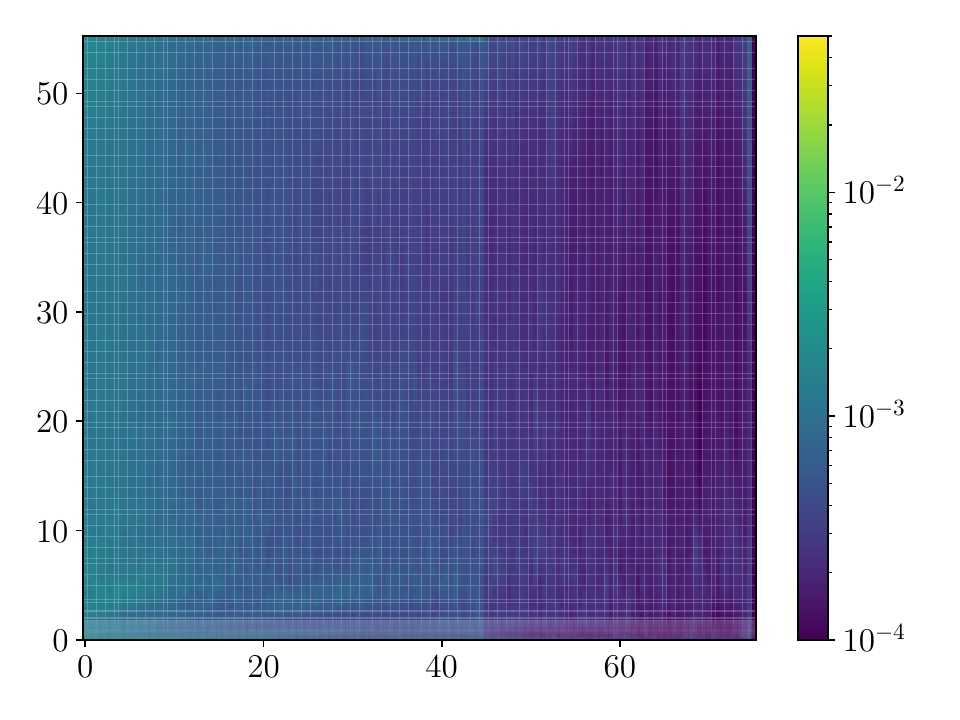}
    \caption{Errors over $A\times X$, averaged over shock values and over distributions from long simulations.}
    \label{fig:app_AggIdioRisk_SpatialErrors}
\end{figure}

To assess the spatial distribution of errors, we compute average errors over long simulations and shock realizations for each grid point $(a_i, x_j)$. Figure~\ref{fig:app_AggIdioRisk_SpatialErrors} displays these errors. Errors are slightly more pronounced for young ages; but importantly, they are not systematically larger around the borrowing constraint. 

\end{document}